\documentclass[runningheads]{llncs}
\usepackage{graphicx}
\usepackage{multirow}
\usepackage{url}
\makeatletter
\g@addto@macro{\UrlBreaks}{\UrlOrds}
\makeatother
\usepackage{multicol}
\usepackage{wrapfig}
\usepackage{xcolor}
\usepackage{hyperref}
\hypersetup{
    colorlinks=true,
    linkcolor=blue,
    citecolor=blue,
    filecolor=blue,      
    urlcolor=blue,
    pdftitle={Towards Effective Paraphrasing for Information Disguise | ECIR 2023},
    pdfpagemode=FullScreen,
    }

\usepackage{enumitem}
\newlist{steps}{enumerate}{1}
\setlist[steps, 1]{label = Step \arabic*:}

\newcommand*\samethanks[1][\value{footnote}]{\footnotemark[#1]}
    \makeatletter
\renewcommand*{\@fnsymbol}[1]{\ensuremath{\ifcase#1\or *\or \dagger\or \ddagger\or
   \mathsection\or \mathparagraph\or \|\or **\or \dagger\dagger
   \or \ddagger\ddagger \else\@ctrerr\fi}}
\makeatother

\begin{document}
\title{Towards Effective Paraphrasing for Information Disguise}

\author{Anmol Agarwal\inst{1}\thanks{\;Authors contributed equally}\orcidID{0000-0001-6730-3565} \and
Shrey Gupta\inst{1}\samethanks\orcidID{0000-0002-5160-2226} \and
Vamshi Bonagiri\inst{1}\orcidID{0000-0002-5537-1664} \and
Manas Gaur\inst{2}\thanks{\;Research with KAI$^2$ Lab @ UMBC}\orcidID{0000-0002-5411-2230} \and 
Joseph Reagle\inst{3}\orcidID{0000-0003-0650-9097} \and \\
Ponnurangam Kumaraguru\inst{1}\orcidID{0000-0001-5082-2078}}
\authorrunning{Agarwal et al.}
%
\institute{IIIT Hyderabad, India \\
\email{\{anmol.agarwal, shrey.gupta\}@students.iiit.ac.in} \\
\email{vamshi.b@research.iiit.ac.in, pk.guru@iiit.ac.in} \and
University of Maryland, Baltimore County, USA \\ 
\email{manas@umbc.edu} \and 
Northeastern University, USA \\
\email{joseph@reagle.org}}
\maketitle              

\begin{abstract}
Information Disguise (\textit{ID}), a part of computational ethics in Natural Language Processing (\textit{NLP}), is concerned with best practices of textual paraphrasing to prevent the non-consensual use of authors’ posts on the Internet. 
Research on ID becomes important when authors’ written online communication pertains to sensitive domains, e.g., mental health. 
Over time, researchers have utilized AI-based automated word spinners (e.g., SpinRewriter, WordAI) for paraphrasing content. 
However, these tools fail to satisfy the purpose of ID as their paraphrased content still leads to the source when queried on search engines. 
There is limited prior work on judging the effectiveness of paraphrasing methods for ID on search engines or their proxies, neural retriever (\textit{NeurIR}) models. 
We propose a framework where, for a given sentence from an author’s post, we perform iterative perturbation on the sentence in the direction of paraphrasing with an attempt to confuse the search mechanism of a NeurIR system when the sentence is queried on it. 
Our experiments involve the subreddit ``r/AmItheAsshole'' as the source of public content and Dense Passage Retriever as a NeurIR system-based proxy for search engines. 
Our work introduces a novel method of phrase-importance rankings using perplexity scores and involves multi-level phrase substitutions via beam search. Our multi-phrase substitution scheme succeeds in disguising sentences 82\% of the time and hence takes an essential step towards enabling researchers to disguise sensitive content effectively before making it public. We also release the code of our approach.\footnote{
\url{https://github.com/idecir/idecir-Towards-Effective-Paraphrasing-for-Information-Disguise}}

\end{abstract}
\section{Introduction}
When a researcher quotes, verbatim, from an online post about a sensitive topic (e.g., politics, mental health, drug use), this could bring additional, unwanted scrutiny to the author of that post \cite{doi:10.1080/13645579.2022.2111816}. The supportive role of Reddit also allows a swarm of tracking technologies to pick the authors of the posts as subjects without consent, which can lead others to authors' profiles and posting history, using which other aspects of personal identity might be inferred \cite{hrw_2022}.

Consequently, some researchers alter verbatim phrases, so their sources are not easily locatable via search services (e.g., Google Search). Researchers leverage traditional (e.g., summarization) or AI-based paraphrasing (e.g., Quillbot \cite{fitria2021quillbot}) methods to disguise the content. These strategies are inspired by Bruckman et al.'s two most prominent methods of disguise: (a) Verbatim Quoting and (b) Paraphrasing \cite{Bruckman2002saa}. Until recently, there have been no quantitative tests of the efficacy of such disguise methods and no description of how to do it well.
In 2022, in an analysis of $19$ Reddit research reports which had claimed to have heavily disguised the content, it was found that $11$ out of the $19$ reports failed to disguise their sources sufficiently; that is, one or more of their sources could be located via search services \cite{Reagle2022drs}. A complementary report \cite{ReagleGaur2022swd} tested the efficacy of both human and automated paraphrasing techniques (i.e., Spin Rewriter and WordAI). The report's authors concluded that while word spinners (typically used for generating plagiarism and content farms) could improve the practice of ethical disguise, the research community needed openly specified techniques whose (non)locatability and fidelity to meaning and fluency were well understood.

In this work, we examine the ID problem through the purview of Black Box Adversarial NLP (e.g., perturbations) and NeurIR. So far, most of the methods in Adversarial NLP focus on downstream tasks such as classification and use labeled datasets to train their adversarial paraphrasing model in a supervised fashion \cite{iyyer-etal-2018-adversarial}. In the context of ID, an effective paraphrase of a sentence should make a semantic NeurIR under-rank the source of the sentence. Our proposed method is entirely \textit{unsupervised} as we use only the document ranks returned by the retriever to guide our model. Our research is not directed toward plagiarism. Instead, it focuses on preventing authors' from being a target of non-consensual experiments because of their online content.

We make the following contributions to the current research: (a) We devise a computational method based on expert rules \cite{ReagleGaur2022swd} that attack phrases using BERT and counter-fitting vectors. (b) We automate the method to prioritize attack locations using perplexity metric to determine phrase importance ranking in Section~\ref{levelOneAttack}. (c) We define a novel adaptation of beam search to make multi-level and multi-word perturbations for dynamic paraphrasing in Section~\ref{multi-levelAttack}. (d) We analyze the success of our proposed approach in Section~\ref{evaluation} as a trade-off between locatability\footnote{A source document has high locatability if a system engine retrieves it in the top-$K$ results when queried with one of the sentences within the document.} and semantic similarity. In addition, experts in journalism and communication studies validate our insights. We use \textit{Universal Sentence Encoder (USE)} \cite{cer2018universal} semantic similarity metric to ensure that the meaning is preserved after paraphrasing. Due to the API limitations on Google Search, we test our approach on a NeurIR system - \textit{Dense Passage Retriever (DPR)} \cite{karpukhin-etal-2020-dense}. 

\section{Related Work}

Paraphrasing is a well-studied problem in NLP literature \cite{zhou2021paraphrase} and so are the limitations of deep language models in NLP \cite{ribeiro-etal-2020-beyond}. However, we study paraphrasing from the perspective of ID, a requirement for ethical research in NLP.
Prior works in Adversarial NLP, such as the work by Alzantot et al., focused on word-level perturbations in sentences that fool a sentiment classifier \cite{alzantot-etal-2018-generating}. Following it, Jia et al. \cite{jia-etal-2019-certified} proposed a family of functions to induce robustness in NLP models working on sentiment classification (\textit{SC}) and natural language inference (\textit{NLI}). Jin et al. \cite{jin2020bert} introduced TextFooler, capable of generating paraphrased text, successfully confusing models for SC and NLI. Experiments with diversity-aware paraphrasing metrics like Jeffrey's divergence, Word Mover's Distance (\textit{WMD}), etc., did not yield sentences that would make the author's identity on search engines non-locatable \cite{xu2018d,zhao-etal-2019-moverscore}. Most prior work on paraphrasing methods are specific to classification \cite{Gao2018BlackBoxGO,garg-ramakrishnan-2020-bae,DBLP:conf/ndss/LiJDLW19,li-etal-2020-bert-attack,Ren2019GeneratingNL,yoo-qi-2021-towards-improving} and NLI \cite{alzantot-etal-2018-generating,Minervini2018AdversariallyRN} tasks. 
For a (query, document) pair, the prior work on adversarial retrieval \cite{raval2020one,10.1145/3539813.3545122,10.1145/3576923} focuses mainly on causing highly relevant/non-relevant documents to be demoted/promoted in the rankings by making minimal changes in the document text. Our work is different in that (i) the document store for the retriever is fixed; (ii) we try to perturb the text in queries to demote the rank of the source post; (iii) we aim to use the paraphrased queries to recreate the paraphrased version of the document which can be made public by researchers. 
Hence, we investigate adversarial retrieval from the Information Disguise perspective and develop a method that performs retriever-guided paraphrasing of content for applications requiring ID. 

\section{Methodology}
\noindent \textbf{Dataset:} We collected $2000$ posts from \textit{r/AmItheAsshole}. Given a post, we split its content into \textit{chunks} (\textit{documents}) with sentence boundaries preserved and include each chunk in the document store for DPR. We extracted $1748$ one-line sentences (averaging $23$ words/sentence) across these posts, which caused at least one of the documents extracted from the source post to be ranked within the top $2$ when queried on DPR.

\noindent \textbf{Problem Formulation:} Given a sentence $s_t$ derived from post $P$, let $\mathcal{R}(s_t, P)$ denote the numerically lowest rank among the ranks of documents derived from post $P$ when $s_t$ is queried on DPR.
We aim to generate $s_p$, i.e., a paraphrase of $s_t$, to maximize $\mathcal{R}(s_p, P)$ for making the post $P$ non-locatable, under the constraints that $Sim(s_t,s_p)$$\ge$ $\epsilon$, where $\epsilon$ is a chosen semantic similarity threshold.

\noindent \textbf{System Architecture:}
Let $s_t$=$\langle$ $w_1$,..,$w_L$ $\rangle$ be a sentence with $L$ tokens. As pointed out in \cite{yoo-qi-2021-towards-improving}, if we intend to perturb $n$ disjoint substrings (phrases) of a string, with each substring having $m$ potential replacements, the search space of potential paraphrases of $s_t$ will have $(m+1)^n-1$ possibilities. To reduce this complexity, we first discuss our method to paraphrase with single-phrase perturbations and then extend it for multi-phrase perturbations using beam search over the search space.

\subsection{Level One Attack with Single phrase perturbation}
\label{levelOneAttack}
\noindent \textbf{Identifying attackable phrases of the sentence:}
For $s_t$, there are $\frac{L*(L-1)}{2} + L$ candidate phrases of the form: $\mathit{ph}\equiv$$\langle$ $w_l$,.., $w_r$ $\rangle$ ($l$$\le$$r$)  which can be replaced. To consider attacking only those substrings with proper independent meaning, we consider only those substrings which are present as a node in the constituency-based parse tree of the query $s_t$ obtained using the Berkeley Neural Parser \cite{kitaev2019multilingual}.\\ 

\begin{figure}[h]
\centering
\includegraphics[width=\textwidth]{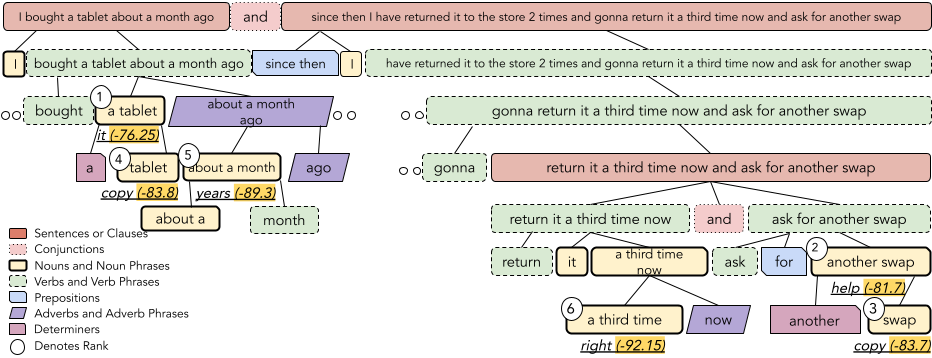}
\caption{Constituency-based parse tree of $s_t$ = ``I bought a tablet about a month ago and since then I have returned it to the store 2 times and gonna return it a third time now and ask for another swap.'' The highest ranked node has a PLL score of $-76$ after replacing \textit{``a tablet''} with \textit{``it''}. The second highest ranked node has a PLL score of $-81.7$ after replacing \textit{``another swap''} with \textit{``help''}.\protect\footnotemark}
\label{fig:1}
\end{figure}

\noindent \textbf{Ranking attackable parse tree nodes based on perplexity:}
    Let the set of nodes in constituency-based parse tree $T$ for $s_t$ be $H$. Let \emph{$N_T^{str}$ be the substring present at node $N_T$} of the parse tree. We define the score PLL($N_T$) for the node as follows:
    \textbf{(Step 1)} Mask $N_T^{str}$ within $s_t$ and use BERT masked language model to find the most likely substitution $Z$.
    \textbf{(Step 2)} Replace mask with $Z$ resulting in new sentence $S$ = $\langle$ $w_1$,..,$w_{l-1}$,Z,$w_{r+1}$,..,$w_L$ $\rangle$.
    \textbf{(Step 3)} $PLL(N_T)$ = Pseudo log-likelihood of the sentence $S$ obtained from BERT by iteratively masking every word in the sentence and then summing the log probabilities, as also done in \cite{alikaniotis-raheja-2019-unreasonable,salazar-etal-2020-masked}.\\
    Let {$AN(s_t)$} = \{${N_T}_1$,.., ${N_T}_P$\} be the top $P$ nodes when ranked on basis of highest $PLL(N_T)$ scores. Since $PLL(s_t)$ is constant across all parse tree nodes, the difference $(PLL(N_T) - PLL(s_t))$ and hence, $PLL(N_T)$ helps us capture the peculiarity of phrase $N_T^{str}$ and hence, its contribution in making the source document $d$ locatable when queried using $s_t$ (see Figure~\ref{fig:1}).\\

\noindent \textbf{Generating suggestions for attacking at a parse tree node:}
For each parse tree node $N_T$ in $AN(s_t)$, we generate candidate perturbations in 2 ways: \textbf{(a)} Bert-masking based candidates $B_{cand}(N_T)$: Generated by masking $N_T^{str}$ within $s_t$ and using BERT to generate $10$ replacements \cite{garg-ramakrishnan-2020-bae}, \textbf{(b)} Synonym-based candidates $CF_{cand}(N_T)$: For nodes containing a single token ($l$$=$$r$), we replace $N_T^{str}$ i.e., $\langle$ $w_{l=r}$ $\rangle$ with the $10$ nearest neighbours in counter-fitting word embedding space \cite{mrksic-etal-2016-counter}, as also done in \cite{yoo-qi-2021-towards-improving}, for producing synonym-based replacements. \\
This leads to the set of candidate perturbations when attacking via the parse tree node $N_T$ to be $Sug_{cand}(N_T) = B_{cand}(N_T)\:{\displaystyle \cup }\:CF_{cand}(N_T) $. As a result, the set of the candidate perturbations (i.e., \textbf{$CP(s_t)$}) derived for query $s_t$ will be the union of candidate perturbations across top-ranked $P$ parse tree nodes in the constituency-based parse tree of $s_t$: \textbf{$CP(s_t)$}  = $\bigcup\limits_{N_T \in AN(s_t)} Sug_{cand}(N_T)$.

\begin{wrapfigure}[18]{R}{0.6\textwidth}
  \begin{center}
    \includegraphics[width=0.6\textwidth, trim={0cm, 0cm, 0cm, 5cm}]{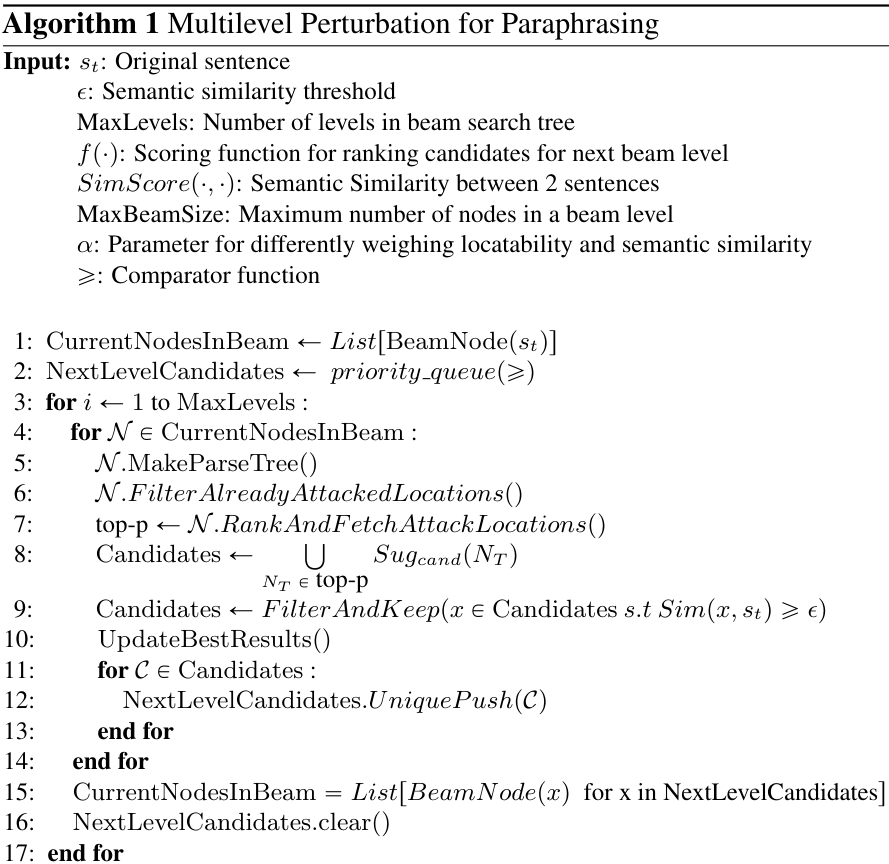}
  \end{center}
  \label{fig:algorithm}
\end{wrapfigure}

\footnotetext{The ethical import of this sample sentence is minimal; it was taken from an innocuous submission (i.e., dings on a tablet) without any identifying information; it was quickly severed from its author (deleted) and consequently does not appear to be indexed by Pushshift or Google. This is applicable for both Figure~\ref{fig:1} and Figure~\ref{fig:2}\label{ethical}}

\subsection{Augmenting the level one attack to multiple levels}
\label{multi-levelAttack}
\begin{figure}
    \centering
    \includegraphics[width=\textwidth]{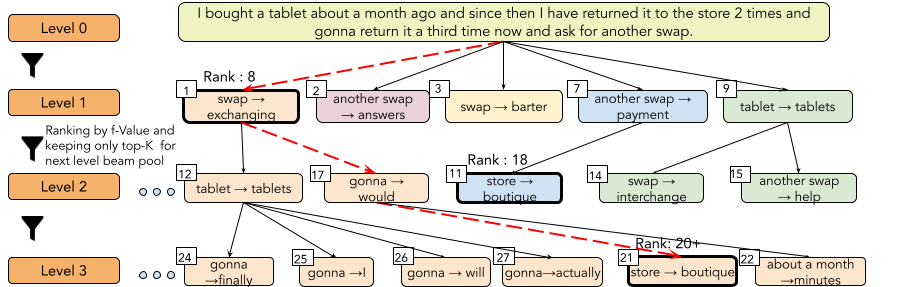}
    \caption{Beam Search Tree for a sample query $s_t$ (only 5 nodes shown at each level for clarity). The best attack at levels $1$ and $2$, corresponding to node IDs $1$ and $11$, pushed $\mathcal{R}(s_p, P)$ from $1$ to $8$ and $18$, respectively. The best attack at level $3$ (shown by the dashed path): “I bought a tablet about a month ago, and since then, I have returned it to the \textit{boutique} 2 times, and \textit{would} return it a third time now and ask for another \textit{exchanging}.”: succeeded in displacing all documents from the source post outside top $20$ while maintaining high semantic similarity of $0.93$ with the original query.}
    \label{fig:2}
\end{figure}

The state-space for multi-phrase perturbation can be considered a tree where the node at level $num$ includes $num$ phrase substitutions on $s_t$ and is obtained using the method described in Section~\ref{levelOneAttack} $num$ times sequentially. As discussed, the size of this search space is vast. Hence, we use \textit{beam search} where the number of nodes expanded at each level in the search tree is restricted to beam width $k$. This selection of $k$ nodes is achieved based on a heuristic function that scores each node based on its potential to have a quality solution in its subtree (see Figure~\ref{fig:2}).

\noindent \textbf{Algorithm Explanations:}
\textbf{(1)} \textit{FilterAlreadyAttackedLocations()}:
Removes those nodes in the parse tree whose phrase has already been replaced once in an attack on one of the previous beam levels. \looseness=-1
\textbf{(2)} \textit{RankAndFetchAttackLocations()}:
Ranks the remaining parse tree nodes based on perplexity scores defined in Section~\ref{levelOneAttack}. \looseness=-1
\textbf{(3)} \textit{f(s)}:
For a candidate $s$, the heuristic score to estimate its potential is calculated as $(1-\alpha) * Sim(s, s_t) + \alpha * (R(s, P)-1)/20$. \looseness=-1
\textbf{(4)} \textit{UniquePush()}:
The priority queue order is determined by operator $\ge$ on $f(s)$ for each candidate $s$. The size is restricted to ``beam width'', and $s$ is not pushed if another element in the queue already has the same text as $s$\looseness=-1 .

\section{Evaluation}
\label{evaluation}
We measure the success of our attack on $s_t$, i.e., query $Q$ based on whether all documents extracted from source post $P$ are absent from the top-K (K=$1$, $5$, $10$, and $20$) retrieved documents by DPR when $s_t$ is queried. To measure the overall effectiveness of our perturbation mechanism, we define the \textit{Hit-Rate@K (HR)} metric given by: \begin{math} HR@K = \frac{(\Sigma_i^{N_q} checkTopK(Q_i))}{N_q} \end{math}, where $checkTopK(Q_i)$ returns $1$ or $0$ depending on whether any of the extracted documents from the target post for query $Q_i$ is in the top-$K$ retrieved documents by DPR and $N_q$ is the total number of queries, i.e., $1748$.  

\begin{table}[!ht]
    \caption{Hit-Rate@K (HR) scores for single-level perturbation approach when attacking using (1) Bert-masking based candidates only, (2) Synonym-based candidates only. Scores are shown across varying values of $\epsilon$ (minimum semantic similarity required between the paraphrase and original sentence) and $P$ (the number of parse tree nodes to attack after perplexity-based ranking).}
    \label{tab:level1Results}
    \centering
        \scriptsize 
    \begin{tabular}{|p{0.27in}|c|c|c|c|c|c|c|c|c|c|c|c|c|c|c|c|c|c|}
\hline
\multirow{3}{*}{HR} & \multicolumn{9}{c|}{Bert-masking based candidates only}&\multicolumn{9}{c|}{Synonym-based candidates only}\\
\cline{2-19}
&\multicolumn{3}{c|}{$\epsilon$=0}&\multicolumn{3}{c|}{$\epsilon$=0.5}&\multicolumn{3}{c|}{$\epsilon$=0.95}&\multicolumn{3}{c|}{$\epsilon$=0}&\multicolumn{3}{c|}{$\epsilon$=0.5}&\multicolumn{3}{c|}{$\epsilon$=0.95}\\
\cline{2-19}
&P=1&P=5&P=*&P=1&P=5&P=*&P=1&P=5&P=*&P=1&P=5&P=*&P=1&P=5&P=*&P=1&P=5&P=*\\
\hline
K=1&0.24&0.07&0.04&0.35&0.09&0.05&0.86&0.56&0.36&0.39&0.19&0.14&0.39&0.20&0.14&0.70&0.40&0.31\\
\hline
K=5&0.48&0.25&0.17&0.59&0.29&0.20&0.96&0.85&0.72&0.70&0.47&0.39&0.70&0.53&0.39&0.91&0.77&0.63\\
\hline
K=10&0.56&0.34&0.26&0.67&0.39&0.29&0.97&0.92&0.83&0.79&0.60&0.52&0.79&0.60&0.52&0.96&0.85&0.76\\
\hline
K=20&0.64&0.44&0.35&0.75&0.50&0.40&0.99&0.96&0.92&0.86&0.71&0.64&0.86&0.71&0.64&0.97&0.92&0.87\\
\hline
    \end{tabular}
\end{table}

For single-level perturbations, as shown in Table~\ref{tab:level1Results}, we compare the performance of the attack schemes using BERT and counter-fitting vectors by varying the semantic similarity threshold ($\epsilon$) and the number of parse tree nodes to be attacked ($P$=$1$, $5$, All/*), after ranking based on perplexity scores. We see that for lower values of $\epsilon$, BERT substitutions are effective at reducing $HR@K$, and preserving the grammatical structure, but the replacements do not take semantics of the replaced phrase into account. Therefore, we filter out suggestions using a high semantic similarity threshold ($\epsilon=0.95$). On the other hand, attacking using counter-fitting vectors replaces words with close synonyms and preserve meaning, so it performs slightly better than BERT on high thresholds, fooling the retriever 23\% of the time when K=5, P=5, and $\epsilon$=0.95 compared to BERT which succeeds only 15\% of the time.

\begin{wraptable}[9]{r}{5.5cm}
    \caption{Hit-Rate@K for Beam Search where: MaxLevels=3, MaxBeamSize=10, $\alpha$=0.8, $\epsilon$=0.8, and P=4.\\}
    \centering
        \scriptsize
    \begin{tabular}{|l|c|c|c|}
    \hline
HR@K & Level 1 & Level 2 &  Level 3\\
    \hline
K=1 & 0.18 & 0.06 & \textbf{0.04} \\
        \hline
K=5 & 0.46 & 0.17 & \textbf{0.10} \\
        \hline
K=10 & 0.60 & 0.24 & \textbf{0.13} \\
        \hline
K=20 & 0.71 & 0.34 & \textbf{0.18} \\
        \hline
    \end{tabular}
 \label{tab:beamResults}
\end{wraptable}

For multi-level attacks using beam search, we use a combination of BERT and counter-fitting based attack strategy owing to the insights obtained from single-level attack experiments. In Table~\ref{tab:beamResults}, we report the $HR@K$ value for \textit{MaxLevels}=\{$1$,$2$,$3$\} i.e., when perturbing \textit{MaxLevels} disjoint phrases within the sentence. We see that only 29\% ($1 - HR@20$ at Level 1) of the attacks work in the first level. The lowest hit rate is obtained at level 3, where attacking just via $4$ parse tree nodes is enough for our attack to succeed 82\% of the time in sending all documents from the source post outside top $20$ despite being constrained under $\epsilon=0.8$. This success rate is higher than the single-level perturbation success rate: 65\% even when $\epsilon=0$ and P=*.
\section{Conclusion}
We introduce a novel black-box framework for effectively paraphrasing text for Information Disguise. Our method uses an unsupervised approach for paraphrasing, where we rank potential attack areas via perplexity scores and generate perturbations using BERT and counter-fitting word vectors. We expand our approach into a multi-phrase substitution setting enabled via beam search. We succeeded in effectively disguising 82\% of the queries by displacing their sources outside a rank of the top 20 when queried on DPR while maintaining high semantic similarity. Our approach can be used to effectively disguise an entire post by concatenating the perturbed versions of the individual sentences within the post from neural retrievers.
However, currently, our approach does not take the grammatical quality of the paraphrased sentences into account. 
Due to the large number of requests made to the retriever to disguise a sentence, our approach is unlikely to work on actual search engines due to API limits. Achieving comparable results as ours while not exceeding the API limits is an interesting problem we wish to solve in our future work.

\bibliography{custom}

\begin{thebibliography}{10}
\providecommand{\url}[1]{\texttt{#1}}
\providecommand{\urlprefix}{URL }
\providecommand{\doi}[1]{https://doi.org/#1}

\bibitem{doi:10.1080/13645579.2022.2111816}
Adams, N.N.: ‘scraping’ reddit posts for academic research? addressing some
  blurred lines of consent in growing internet-based research trend during the
  time of covid-19. International Journal of Social Research Methodology
  \textbf{0}(0),  1--16 (2022). \doi{10.1080/13645579.2022.2111816},
  \url{https://doi.org/10.1080/13645579.2022.2111816}

\bibitem{alikaniotis-raheja-2019-unreasonable}
Alikaniotis, D., Raheja, V.: The unreasonable effectiveness of transformer
  language models in grammatical error correction. In: Proceedings of the
  Fourteenth Workshop on Innovative Use of NLP for Building Educational
  Applications. pp. 127--133. Association for Computational Linguistics,
  Florence, Italy (Aug 2019). \doi{10.18653/v1/W19-4412},
  \url{https://aclanthology.org/W19-4412}

\bibitem{alzantot-etal-2018-generating}
Alzantot, M., Sharma, Y., Elgohary, A., Ho, B.J., Srivastava, M., Chang, K.W.:
  Generating natural language adversarial examples. In: Proceedings of the 2018
  Conference on Empirical Methods in Natural Language Processing. pp.
  2890--2896. Association for Computational Linguistics, Brussels, Belgium
  (Oct-Nov 2018). \doi{10.18653/v1/D18-1316},
  \url{https://aclanthology.org/D18-1316}

\bibitem{Bruckman2002saa}
Bruckman, A.: Studying the amateur artist: A perspective on disguising data
  collected in human subjects research on the {Internet}. Ethics and
  Information Technology  \textbf{4}(3) (2002),
  \url{http://citeseerx.ist.psu.edu/viewdoc/download?doi=10.1.1.432.1591&rep=rep1&type=pdf}

\bibitem{cer2018universal}
Cer, D., Yang, Y., Kong, S.y., Hua, N., Limtiaco, N., John, R., Constant, N.,
  Guajardo-Cespedes, M., Yuan, S., Tar, C., Sung, Y.H., Strope, B., Kurzweil,
  R.: Universal sentence encoder  (03 2018)

\bibitem{fitria2021quillbot}
Fitria, T.N.: Quillbot as an online tool: Students’ alternative in
  paraphrasing and rewriting of english writing. Englisia: Journal of Language,
  Education, and Humanities  \textbf{9}(1),  183--196 (2021)

\bibitem{Gao2018BlackBoxGO}
Gao, J., Lanchantin, J., Soffa, M.L., Qi, Y.: Black-box generation of
  adversarial text sequences to evade deep learning classifiers. 2018 IEEE
  Security and Privacy Workshops (SPW) pp. 50--56 (2018)

\bibitem{garg-ramakrishnan-2020-bae}
Garg, S., Ramakrishnan, G.: {BAE}: {BERT}-based adversarial examples for text
  classification. In: Proceedings of the 2020 Conference on Empirical Methods
  in Natural Language Processing (EMNLP). pp. 6174--6181. Association for
  Computational Linguistics, Online (Nov 2020).
  \doi{10.18653/v1/2020.emnlp-main.498},
  \url{https://aclanthology.org/2020.emnlp-main.498}

\bibitem{hrw_2022}
HRW: ``how dare they peep into my private life?" (Oct 2022),
  \url{https://www.hrw.org/report/2022/05/25/how-dare-they-peep-my-private-life/childrens-rights-violations-governments}

\bibitem{iyyer-etal-2018-adversarial}
Iyyer, M., Wieting, J., Gimpel, K., Zettlemoyer, L.: Adversarial example
  generation with syntactically controlled paraphrase networks. In: Proceedings
  of the 2018 Conference of the North {A}merican Chapter of the Association for
  Computational Linguistics: Human Language Technologies, Volume 1 (Long
  Papers). pp. 1875--1885. Association for Computational Linguistics, New
  Orleans, Louisiana (Jun 2018). \doi{10.18653/v1/N18-1170},
  \url{https://aclanthology.org/N18-1170}

\bibitem{jia-etal-2019-certified}
Jia, R., Raghunathan, A., G{\"o}ksel, K., Liang, P.: Certified robustness to
  adversarial word substitutions. In: Proceedings of the 2019 Conference on
  Empirical Methods in Natural Language Processing and the 9th International
  Joint Conference on Natural Language Processing (EMNLP-IJCNLP). pp.
  4129--4142. Association for Computational Linguistics, Hong Kong, China (Nov
  2019). \doi{10.18653/v1/D19-1423}, \url{https://aclanthology.org/D19-1423}

\bibitem{jin2020bert}
Jin, D., Jin, Z., Zhou, J.T., Szolovits, P.: Is bert really robust? a strong
  baseline for natural language attack on text classification and entailment.
  In: Proceedings of the AAAI conference on artificial intelligence. vol.~34,
  pp. 8018--8025 (2020)

\bibitem{karpukhin-etal-2020-dense}
Karpukhin, V., Oguz, B., Min, S., Lewis, P., Wu, L., Edunov, S., Chen, D., Yih,
  W.t.: Dense passage retrieval for open-domain question answering. In:
  Proceedings of the 2020 Conference on Empirical Methods in Natural Language
  Processing (EMNLP). pp. 6769--6781. Association for Computational
  Linguistics, Online (Nov 2020). \doi{10.18653/v1/2020.emnlp-main.550},
  \url{https://aclanthology.org/2020.emnlp-main.550}

\bibitem{kitaev2019multilingual}
Kitaev, N., Cao, S., Klein, D.: Multilingual constituency parsing with
  self-attention and pre-training. In: Proceedings of the 57th Annual Meeting
  of the Association for Computational Linguistics. pp. 3499--3505 (2019)

\bibitem{DBLP:conf/ndss/LiJDLW19}
Li, J., Ji, S., Du, T., Li, B., Wang, T.: Textbugger: Generating adversarial
  text against real-world applications. In: 26th Annual Network and Distributed
  System Security Symposium, {NDSS} 2019, San Diego, California, USA, February
  24-27, 2019. The Internet Society (2019),
  \url{https://www.ndss-symposium.org/ndss-paper/textbugger-generating-adversarial-text-against-real-world-applications/}

\bibitem{li-etal-2020-bert-attack}
Li, L., Ma, R., Guo, Q., Xue, X., Qiu, X.: {BERT}-{ATTACK}: Adversarial attack
  against {BERT} using {BERT}. In: Proceedings of the 2020 Conference on
  Empirical Methods in Natural Language Processing (EMNLP). pp. 6193--6202.
  Association for Computational Linguistics, Online (Nov 2020).
  \doi{10.18653/v1/2020.emnlp-main.500},
  \url{https://aclanthology.org/2020.emnlp-main.500}

\bibitem{Minervini2018AdversariallyRN}
Minervini, P., Riedel, S.: Adversarially regularising neural nli models to
  integrate logical background knowledge. In: Conference on Computational
  Natural Language Learning (2018)

\bibitem{mrksic-etal-2016-counter}
Mrk{\v{s}}i{\'c}, N., {\'O}~S{\'e}aghdha, D., Thomson, B., Ga{\v{s}}i{\'c}, M.,
  Rojas-Barahona, L.M., Su, P.H., Vandyke, D., Wen, T.H., Young, S.:
  Counter-fitting word vectors to linguistic constraints. In: Proceedings of
  the 2016 Conference of the North {A}merican Chapter of the Association for
  Computational Linguistics: Human Language Technologies. pp. 142--148.
  Association for Computational Linguistics, San Diego, California (Jun 2016).
  \doi{10.18653/v1/N16-1018}, \url{https://aclanthology.org/N16-1018}

\bibitem{raval2020one}
Raval, N., Verma, M.: One word at a time: adversarial attacks on retrieval
  models. arXiv preprint arXiv:2008.02197  (2020)

\bibitem{Reagle2022drs}
Reagle, J.: Disguising {Reddit} sources and the efficacy of ethical research.
  Ethics and Information Technology  \textbf{24}(3) (sep 2022),
  \url{http://dx.doi.org/10.1007/s10676-022-09663-w}

\bibitem{ReagleGaur2022swd}
Reagle, J., Gaur, M.: Spinning words as disguise: {Shady} services for ethical
  research? First Monday  (jan 2022),
  \url{http://dx.doi.org/10.5210/fm.v27i1.12350}

\bibitem{Ren2019GeneratingNL}
Ren, S., Deng, Y., He, K., Che, W.: Generating natural language adversarial
  examples through probability weighted word saliency. In: Annual Meeting of
  the Association for Computational Linguistics (2019)

\bibitem{ribeiro-etal-2020-beyond}
Ribeiro, M.T., Wu, T., Guestrin, C., Singh, S.: Beyond accuracy: Behavioral
  testing of {NLP} models with {C}heck{L}ist. In: Proceedings of the 58th
  Annual Meeting of the Association for Computational Linguistics. pp.
  4902--4912. Association for Computational Linguistics, Online (Jul 2020).
  \doi{10.18653/v1/2020.acl-main.442},
  \url{https://aclanthology.org/2020.acl-main.442}

\bibitem{salazar-etal-2020-masked}
Salazar, J., Liang, D., Nguyen, T.Q., Kirchhoff, K.: Masked language model
  scoring. In: Proceedings of the 58th Annual Meeting of the Association for
  Computational Linguistics. pp. 2699--2712. Association for Computational
  Linguistics, Online (Jul 2020). \doi{10.18653/v1/2020.acl-main.240},
  \url{https://aclanthology.org/2020.acl-main.240}

\bibitem{10.1145/3539813.3545122}
Wang, Y., Lyu, L., Anand, A.: Bert rankers are brittle: A study using
  adversarial document perturbations. In: Proceedings of the 2022 ACM SIGIR
  International Conference on Theory of Information Retrieval. p. 115–120.
  ICTIR '22, Association for Computing Machinery, New York, NY, USA (2022).
  \doi{10.1145/3539813.3545122}, \url{https://doi.org/10.1145/3539813.3545122}

\bibitem{10.1145/3576923}
Wu, C., Zhang, R., Guo, J., de~Rijke, M., Fan, Y., Cheng, X.: Prada: Practical
  black-box adversarial attacks against neural ranking models. ACM Trans. Inf.
  Syst.  (dec 2022). \doi{10.1145/3576923},
  \url{https://doi.org/10.1145/3576923}, just Accepted

\bibitem{xu2018d}
Xu, Q., Zhang, J., Qu, L., Xie, L., Nock, R.: D-page: Diverse paraphrase
  generation. CoRR  \textbf{abs/1808.04364} (2018),
  \url{http://arxiv.org/abs/1808.04364}

\bibitem{yoo-qi-2021-towards-improving}
Yoo, J.Y., Qi, Y.: Towards improving adversarial training of {NLP} models. In:
  Findings of the Association for Computational Linguistics: EMNLP 2021. pp.
  945--956. Association for Computational Linguistics, Punta Cana, Dominican
  Republic (Nov 2021). \doi{10.18653/v1/2021.findings-emnlp.81},
  \url{https://aclanthology.org/2021.findings-emnlp.81}

\bibitem{zhao-etal-2019-moverscore}
Zhao, W., Peyrard, M., Liu, F., Gao, Y., Meyer, C.M., Eger, S.: {M}over{S}core:
  Text generation evaluating with contextualized embeddings and earth mover
  distance. In: Proceedings of the 2019 Conference on Empirical Methods in
  Natural Language Processing and the 9th International Joint Conference on
  Natural Language Processing (EMNLP-IJCNLP). pp. 563--578. Association for
  Computational Linguistics, Hong Kong, China (Nov 2019).
  \doi{10.18653/v1/D19-1053}, \url{https://aclanthology.org/D19-1053}

\bibitem{zhou2021paraphrase}
Zhou, J., Bhat, S.: Paraphrase generation: A survey of the state of the art.
  In: Proceedings of the 2021 Conference on Empirical Methods in Natural
  Language Processing. pp. 5075--5086 (2021)

\end{thebibliography}
\bibliographystyle{splncs04}

\end{document}